\documentclass[pra,twocolumn,superscriptaddress,showpacs,longbibliography]{revtex4-1}
\usepackage{graphicx}
\usepackage{latexsym}
\usepackage{amsmath}
\usepackage{graphics}
\usepackage{amssymb}
\usepackage{layout}
\usepackage{verbatim}
\usepackage{amsfonts,epsfig,dsfont,bbm}
\usepackage{natbib}
\usepackage{hyperref}
\usepackage{color}

\newtheorem{prop}{Proposition}

\DeclareMathOperator{\Tr}{Tr}

\begin{document}

%\preprint{}

\title{Entanglement and  objectivity in pure dephasing models}% Force line breaks with \\

\author{Katarzyna Roszak}
\affiliation{Department of Theoretical Physics, Faculty of Fundamental Problems of Technology, Wroc{\l}aw University of Science and Technology,
50-370 Wroc{\l}aw, Poland}

\author{Jaros\l aw K.~Korbicz}
\affiliation{Center for Theoretical Physics, Polish Academy of Sciences, Aleja Lotników
	32/46, 02-668 Warsaw, Poland}

\date{\today}% It is always \today, today,

\begin{abstract}
	We study the relation between the emergence of objectivity and qubit-environment entanglement
	generation. We find that although entanglement with the unobserved environments is irrelevant
	(since sufficiently strong decoherence can occur regardless), entanglement with the observed
	environments is crucial. In fact, the appearance of an objective qubit-observed-environment state is strictly impossible if their joint evolution does not lead to entanglement. 
	Furthermore, if a single observer has access to a single environment (no macrofractions)
	then the required orthogonality of the observed environmental states comes only as a 
	consequence of the system-enviornment state becoming strongly entangled (maximally
	entangled for the given initial occupation of the qubit if the environmental state
	is initially pure). 
\end{abstract}

\pacs{}% PACS, the Physics and Astronomy
\maketitle

\section{Introduction}
The idea of objectivity \cite{ollivier04,ollivier05,zurek09} is related to the
notion that classicality should emerge naturally out of the quantum description
to allow unrelated observations of a system by different parties,
which would neither destroy the state of the observed system
nor leave ambiguity in the information obtained about its state. 
Such observations are to be performed
not by direct measurements of the system of interest (which would obviously
not satisfy the requirements of objectivity in quantum mechanics), but rather by 
measurements of different environments which can gain information about the system state
during their joint evolution with said system. 

Recently, the notion of such a system-environment state has been has been specified
and its mathematical structure proposed \cite{korbicz14,horodecki15}.
The so called spectrum broadcast structure (SBS) states are zero-discord states
from the point of view of the system of interest and from each environment to be observed,
which guarantees that repeated, independent measurements on individual environments
do not disturb the state of the system nor of other environments:
\begin{eqnarray}\label{SBS_0}
&&\hat\sigma_{SBS}=\sum_ip_i|i\rangle\langle i| \otimes\rho^{1}_i...\otimes\rho^{N}_i,\\
&&\hat\rho^{k}_i\perp \hat\rho^{k}_{i'} \textrm{ for every } i'\ne i \textrm{ and } k={1,\dots,N}.
\end{eqnarray}
Here $\{|i\rangle\}$ is the so-called pointer basis of the central system to which it decoheres, $p_i$ are initial pointer probabilities,
$k$ enumerates the environments, and $\hat\rho^{k}_i$ are some states of the observed parts of the environment of which we require only that they have 
mutually orthogonal supports for different pointer index $i$ or in other words are
perfectly distinguishable in one shot.

To study situations when the emergence of objectivity is possible in the above sense of SBS creation, typically system-environment
interactions are taken into account which lead to pure dephasing of the qubit 
\cite{tuziemski15,mironowicz17,korbicz17,lampo17,korbicz17a,tuziemski18,mironowicz18,tuziemski19}
after the environmental degrees of freedom are 
traced out. 
These types of interactions describe situations when the $T_2^*$ times 
are much shorter than the $T_1$ times (the decay of phase coherence is much faster
than relaxation) are fairly common in solid state scenarios.
It is true that for the two-level systems earliest studied as qubits in context of quantum information processing, such as two-level atoms manipulated coherently with a narrow linewidth laser field, energy exchange with an environment (spontaneous emission) was often the most relevant decoherence mechanism. However, this was due to the fact that the only environment of such atoms was electromagnetic vacuum. In order to advance the development of quantum circuits to the multi-qubit stage, it turned out to be necessary to move to architectures in which the relevant environment was much more structured. All the solid-state based qubits (both spin based 
\cite{medford13,staudacher13,muhonen14,romach15,malinowski17} and charge-based \cite{borri01,vagov04,roszak06a}  in semiconductors, flux \cite{bylander11} and many charge-based qubits \cite{nakamura02a} in superconductors) are exposed to structured environments – lattice vibrations, charge noise from various sources, nuclear spin fluctuations etc. – which typically have slow dynamics (correlation times possibly longer than typical qubit decoherence time), and as such are more efficient at dephasing of the qubit state that is in a superposition of pointer states, than at changing the populations of these pointer states. It is important to note that the same applies to ion trap qubits \cite{biercuk09}. In all these systems the dephasing of the qubit occurs on timescales orders of magnitude shorter than the timescale of energy exchange with the environment, and decoherence channels such as amplitude damping are irrelevant when the initial state of the qubit is a superposition of pointer states. 

Such interactions do not disturb the occupations of the system
described in the basis of so called pointer states \cite{Zurek_PRD81,Zurek_RMP03}, but are 
detrimental to qubit coherence.
When a system-environment state is initially in a pure
(product) state, the decoherence corresponds directly to the buildup of system-environment entanglement,
and is in fact accompanied by the transfer of information about the system state 
into the environment. This information transfer is the basis for the possible formation
of SBS states, since for information about the system state to be read out of an environment,
it must first be transferred there by some physical process. The formation of SBS states
further requires a source of decoherence (and environment or environments which are not
observed) as to lose the coherence and correlations between the system and the observed 
environments, which is necessary to get the zero-discord form.

In realistic situations, pure initial states of any environments are rare. Hence, the study
of mixed initial environmental states is necessary to assess if SBS states are likely to 
occur. For mixed states the correlation between system-environment entanglement buildup
and system decoherence is less direct; although such entanglement is always accompanied by 
decoherence, decoherence is also possible without said entanglement \cite{Eisert_PRL02,Hilt_PRA09,Pernice_PRA11,roszak15a,roszak18}.
In the following we use the methods devised in Refs \cite{roszak15a,roszak18}
to study the interrelations between system-environment entanglement generation
and the possibility of the emergence of SBS states. We show that entanglement is a necessary,
but not sufficient condition for objectivity. 
In fact, if no entanglement between the observed environments and the system
is generated, there is no possibility to distinguish between the system pointer states
by performing any measurements on the environments.
Hence, entanglement is directly responsible for 
the transfer of information about the system state into an environment
and its lack is equivalent to the complete absence of such transfer,
so no methods which can be used to enhance distinguishability 
(such as assigning more environments to a single observer 
\cite{tuziemski15,mironowicz17,korbicz17,lampo17,korbicz17a,tuziemski18,mironowicz18,tuziemski19})
will work.

We further study the conditions 
which need to be fulfilled by the interaction with the observed environments 
for environmental states corresponding to qubit
pointer states to become fully distinguishable (this part is limited
to the system consisting of a single qubit).
We find that the amount of entanglement necessary is surprisingly large 
and that, consequently, there are constraints on the initial purity of the environments
for such full distinguishability to be able to manifest itself.

The article is organized as follows. In Sec.~\ref{sec2} we introduce the studied model in 
detail, including the Hamiltonian of the system, observed environments, and unobserved
environments, the resulting evolution operator and the structure of the density matrix
of the whole system at time $t$. We then show the necessary constraints on the evolution
of the observed and unobserved environments conditional on the states of the system
for the emergence of objectivity. In Sec.~\ref{sec3} we first show that separability of
the qubit and its observed environments excludes the emergence of SBS states,
and later generalize the results to a system of any size.
Sec.~(\ref{sec4}) is devoted to the study of when observations of an environment 
can be used to completely distinguish between qubit pointer states. We find that this leads
to a strict condition for qubit-environment entanglement, but also excludes small values
of initial purity of the environment. Sec.~(\ref{sec5}) concludes the article.

\section{Pure dephasing evolutions and SBS states \label{sec2}}

The measurement Hamiltonian which is used to study the emergence
of spectrum broadcast structure states is, in fact
a special case of the general Hamiltonian which leads to pure dephasing evolutions
\cite{tuziemski15,mironowicz17,korbicz17,lampo17,korbicz17a,tuziemski18,mironowicz18,tuziemski19} in a system-environment scenario \cite{roszak15a,roszak18}. 
The Hamiltonian is given by
\begin{equation}
\label{ham}
\hat{H}=\sum_{i}\left(
\varepsilon_i|i\rangle\langle i|
+|i\rangle\langle i|\otimes\hat{V}_{not}^i+
\sum_{k}|i\rangle\langle i|\otimes{\hat{V}_k^i}\right),
\end{equation}
where the first term on the right describes the free Hamiltonian of the system
in its eigenbasis $\{|i\rangle\}$, while the other two
terms describe the interaction with the environment. The environmental
coupling is divided into two parts: The second term on the right
is responsible for the interaction with the part of the environment
which is not observed and its free evolution, $\hat{V}_{not}^i
=\hat{H}_{not}+\hat{\tilde{V}}_{not}^i$. This Hamiltonian may be further
resolved into parts describing different unobserved environments.
We do not do it explicitly here, as it has no bearing on the described
results. The third term is 
responsible for the interaction of the system with the observed
part of the environment and its free evolution, $\hat{V}_{k}^i
=\hat{H}_{k}+\hat{\tilde{V}}_{k}^i$. This part of the Hamiltonian
is resolved with respect to the individual observers, which are labeled
by the index $k$.
Since all terms in the Hamiltonian (\ref{ham}) commute, the evolution
operator for the whole system can be written as
\begin{equation}
\label{evol}
\hat{U}(t)=\sum_ie^{-i\varepsilon_it}|i\rangle\langle i |\otimes
e^{-i\hat{V}_{not}^i t}\otimes\bigotimes_ke^{-i \hat{V}_{k}^it}.
\end{equation}
Let us stress here that the free evolution of the environments is 
present in the environmental coupling operators $\hat{V}_{not}^i$
and $\hat{V}_{k}^i$ and no assumptions are made on the commutation relations
between different parts of the Hamiltonian pertaining to the same environment
(neither the free Hamiltonian must commute with the interaction, nor 
the different elements of the interaction with themselves).
The difference between Hamiltonian (\ref{ham}) and the general pure-dephasing
Hamilotnian amounts to the lack of interaction between the different environments.

We assume (following the papers on objectivity 
\cite{tuziemski15,mironowicz17,korbicz17,lampo17,korbicz17a,tuziemski18,mironowicz18,tuziemski19}
that the initial state of the system and environments is a product
state with respect to the system, the unobserved environment, and
each observed environment. 
This translates into the physical situation, when the state of the system
can be prepared without disturbing any of the environments, and the observers
each are truly separated. 
Furthermore we assume (as to be able
to use the results on system-environment entanglement 
for pure dephasing evolutions \cite{roszak15a,roszak18}) that the initial 
state of the qubit is pure. This means that the initial state
of the system and its environments can be written as
\begin{equation}
\label{ini}
\hat{\tilde{\sigma}}(0)=|\psi_0\rangle\langle\psi_0|\otimes\hat{R}_{not}(0)
\otimes\bigotimes_k\hat{\rho}^k(0),
\end{equation}
where $|\psi_0\rangle=\sum_ia_i|i\rangle$ is a superposition
of pointer states of the system, $\hat{R}_{not}(0)$ is the initial state
of the unobserved environment, and $\hat{\rho}^k(0)$ are the individual initial states
of the observed environments.

Taking the initial state
(\ref{ini}) and using the evolution operator given by eq.~(\ref{evol})
yields the density matrix of the system and its environments
at any time $t$. If the system is only a qubit, $i=0,1$,
this can be explicitly written as 
\begin{widetext}
\begin{equation}
\label{caly}
\hat{\tilde{\sigma}}(t)=\left(
\begin{array}{cc}
|a_0|^2\hat{R}_{00}(t)\otimes\bigotimes_k\hat{\rho}_{00}^k(t)&
a_0a_1^*(t)
\hat{R}_{01}(t)\otimes\bigotimes_k\hat{\rho}_{01}^k(t)\\
a_1(t)a_0^*
\hat{R}_{10}(t)\otimes\bigotimes_k\hat{\rho}_{10}^k(t)&
|a_1|^2
\hat{R}_{11}(t)\otimes\bigotimes_k\hat{\rho}_{11}^k(t)
\end{array} 
\right),
\end{equation}
\end{widetext}
where $a_1(t)=be^{-i\Delta\varepsilon t}$ and $\Delta\varepsilon=\varepsilon_1-\varepsilon_0$,
\begin{subequations}
\begin{eqnarray}
\label{rija}
\hat{R}_{ij}(t)&=&\hat{w}_i^{not}(t)\hat{R}_{not}(0)\hat{w}_j^{not\dagger},\\
\label{rijb}
\hat{\rho}_{ij}^k(t)&=&\hat{w}_i^{k}(t)\hat{\rho}^k(0)\hat{w}_j^{k\dagger}(t),
\end{eqnarray}
\end{subequations}
with the environmental evolution operators conditional on the state of the qubit
given by
\begin{subequations}
	\begin{eqnarray}
	\label{wia}
	\hat{w}_i^{not}(t)&=&e^{-i\hat{V}_{not}^i t},\\
	\label{wib}
	\hat{w}_i^{k}(t)&=&e^{-i\hat{V}_{k}^i t}.
	\end{eqnarray}
\end{subequations}
This structure is preserved, and it can be easily generalized to a system of any size 
\cite{roszak18}.

The next step in trying to obtain the SBS state is tracing out
over the unobserved environment, which for a qubit yields
\begin{equation}
\label{pot}
\hat{\sigma}(t)=\left(
\begin{array}{cc}
|a_0|^2\bigotimes_k\hat{\rho}_{00}^k(t)&
a_0a_1^*(t)
\Gamma_{01}(t)\bigotimes_k\hat{\rho}_{01}^k(t)\\
a_1(t)a_0^*
\Gamma_{01}^*(t)\bigotimes_k\hat{\rho}_{10}^k(t)&
|a_1|^2
\bigotimes_k\hat{\rho}_{11}^k(t)
\end{array} 
\right),
\end{equation}
since $\Tr \hat{R}_{ii}(t)=1$
(the matrices $\hat{R}_{ii}(t)$ are density matrices, as
they are obtained via a unitary operation from the initial
density matrix of the unobserved environment) and
where 
\begin{equation}\label{gamma}
\Gamma_{01}(t)=\Tr \hat{R}_{01}(t)=\Tr\left[\hat{w}_1^{not\dagger}(t)\hat{w}_0^{not}(t)\hat{R}_{not}(0)\right]
\end{equation}
is the decoherence factor.

For the state (\ref{pot}) to become an SBS state in the
 $\{|0\rangle ,|1\rangle\}$ pointer basis of the qubit, two conditions need to be met.
Firstly, the decoherence function stemming from the unobserved part of
the environment has to decay to zero, $\Gamma_{01}(t)=0$. In the following
we will not be studying this condition, since 
its fulfillment is not related to the generation of qubit-environment
entanglement, but only to the strength of the interaction
between the qubit and the unobserved part of the enviornment. In fact, the condition can be met both in case
of entangling and non-entangling evolutions \cite{Eisert_PRL02,Hilt_PRA09,Pernice_PRA11,roszak15a,roszak18}, and we will implicitly
assume that the interaction is strong enough to cancel the
off-diagonal terms in eq.~(\ref{pot}).
Under such an assumption, the density matrix (\ref{pot}) is given
by 
\begin{equation}
\label{SBS}
\hat{\sigma}(t)=\sum_i
|a_i|^2|i\rangle\langle i|\otimes\bigotimes_k\hat{\rho}_{ii}^k(t).
\end{equation}
For larger systems this form is obtained when all of the decoherence factors
$\Gamma_{ij}(t)=\Tr \hat{R}_{ij}(t)=\Tr\left[\hat{w}_i^{not\dagger}(t)\hat{w}_j^{not}(t)\hat{R}_{not}(0)\right]=0$,
for all $i\neq j$.

For the density matrix of the system and observed environments
to be an SBS state, a second requirement is necessary, namely
that the observed environment density matrices conditional on the system
pointer states, $\hat{\rho}_{ii}^k(t)$, be
perfectly distinguishable. This in turn is equivalent to them being orthogonal, 
in the sense of having orthogonal supports
\begin{equation}
\label{warunek}
\hat{\rho}_{ii}^k(t)\hat{\rho}_{jj}^k(t)=0
\end{equation}
for every pair of system pointer states, $i\neq j$, and
for all observed environments $k$.
Only then can the system states be uniquely determined 
by measurements on any of the environments, without damaging
either the system state or the states of the environments themselves \cite{korbicz14,horodecki15}.

\section{Entanglement and the emergence of objectivity\label{sec3}}
Let us begin with the study of the situation when the system of interest 
is only a qubit (so $i=0,1$).
In Ref.~(\cite{roszak15a}) it is shown that for an evolution 
governed by a Hamiltonian which leads to pure dephasing of the qubit
after the environmental degrees of freedom are traced out, of which the Hamiltonian
given in 	
eq.~(\ref{ham}) is a special case, and a product initial qubit-environment state,
with the qubit initially in a pure state, as in the initial state
of eq.~(\ref{ini}), the qubit-environment state is separable 
if and only if the evolutions of the environment conditional
on the qubit being in one of its pointer states are the same.
In the studied scenario, this condition translates into 
\begin{equation}
\label{sep}
\hat{R}_{ii}(t)\otimes\bigotimes_k\hat{\rho}_{ii}^k(t)
=\hat{R}_{jj}(t)\otimes\bigotimes_k\hat{\rho}_{jj}^k(t),
\end{equation}
with $i=0$ and $j=1$.
Since all of the environmental states in eq.~(\ref{sep}) are 
in product form, and the environments themselves are described
by different subspaces of the Hilbert space,
the separability condition can be resolved into separate conditions
for each environment $k$ (the unobserved environment is irrelevant,
since its entanglement is irrelevant with respect to objectivity).  
Hence, if there is no entanglement generated between the qubit
and the environment, then
\begin{equation}
\label{sep2}
\forall_k\;\hat{\rho}_{ii}^k(t)
=\hat{\rho}_{jj}^k(t),
\end{equation}
Obviously 
this implies that the state (\ref{SBS}) cannot be an SBS state, as states $\hat{\rho}_{ii}^k(t)$ are then identical for different central qubit states $i$.

This result can be easily generalized to a system of any size with the help of 
Ref.~(\cite{roszak18}), where it is shown that the necessary condition for separability
(but not sufficient) is exactly of the form as eq.~(\ref{sep}), but
it must be fulfilled for all pairs of system pointer states, $i\neq j$.
This translates into a family of conditions for different environments $k$, as in eq.~(\ref{sep2}).
Hence, for systems larger than a qubit, objectivity cannot emer{ge if the the system-environment
evolution for state $k$ is non-entangling, but also for entangling evolutions
that satisfy the necessary separability condition (\ref{sep2}).

This leads us to our first result:

\begin{prop}
If the evolution governed by a pure dephasing Hamiltonian (\ref{ham}) with an initial state of the form (\ref{ini})
does not generate a qubit-environment entanglement, then SBS states will not be formed. 
\end{prop}

%since
%$\hat{\rho}_{ii}^k(t)$ are density matrices, and it is impossible 
%for $(\hat{\rho}_{ii}^k(t))^2=0$, so the second condition for 
%objectivity (\ref{warunek}) cannot be met, when there is no 
%qubit-environment entanglement generated during the evolution. 

Thus we obtain an interesting result that qubit-environment entanglement (QEE) is a necessary condition for the emergence of
objectivity. This is somewhat intriguing since the purely quantum property of entanglement appears to be 
necessary for the emergence  of the classical property of objectivity (for a  similar conclusion but obtained in a very different context see Ref.~\cite{richens17}).  
But QEE alone is not sufficient
for the emergence of objectivity even for qubit systems, since the condition for QEE generation,
 $\hat{\rho}_{ii}^k(t)\neq \hat{\rho}_{jj}^k(t)$, is much weaker than the condition
(\ref{warunek}).  We study the latter in more detail in the next Section.

%This can be seen, when examining
%the separability condition (\ref{sep2}), which yields the extreme
%opposite of orthogonality between the matrices $\hat{\rho}_{00}^k(t)$
%and $\hat{\rho}_{11}^k(t)$, since they are the same. By extrapolation,
%there must exist states for which $\hat{\rho}_{00}^k(t)\neq \hat{\rho}_{11}^k(t)$ (so QEE is generated during the evolution),
%while $\hat{\rho}_{00}^k(t)\hat{\rho}_{11}^k(t)\neq 0$.

\section{Strict distinguishability\label{sec4}}
Let us now study the situation when the orthogonality condition (\ref{warunek})
is strictly fulfilled. In realistic situations one can hardly expect 
such strict fulfillment (see e.g.~
\cite{tuziemski15,mironowicz17,korbicz17,lampo17,korbicz17a,tuziemski18,mironowicz18,tuziemski19}) and some measure of 
distinguishability \cite{calsamiglia08,fuchs99}, like the state fidelity, must be used. Nevertheless it is interesting
to study the ideal situation and use it to infer about the behavior of systems which do not
show strict orthogonality, but do exhibit the generation of system-environment entanglement. 
The following will be restricted to the system of interest composed of a single qubit.
%If this condition is not strictly fulfilled
%then a single environment is insufficient for the emergence of 
%objectivity, and an SBS state will not be obtained when tracing out
%the unobserved environments. 
%a macrofraction [Korbicz Horodecki^2 PRL 2014] of multiple environments is available to a single
%observer [].

The fulfillment of condition (\ref{warunek}) means that 
the environmental density matrices conditional on different qubit states
$|0\rangle$ and $|1\rangle$ must be defined on
separate subspaces of the Hilbert space of the corresponding environment.
Hence, it imposes strong limitations
on the evolution of the qubit-environment state. 
Since the condition (\ref{warunek}) must be fulfilled separately
for any environment $k$, the study of the condition for a single environment 
leads to results which must be fulfilled by all observed environments separately.
Hence, we will study one environment labeled by the index $k$
without any limitations
on its dimension $d_k$. 

It is now most convenient to express all of the $\hat{\rho}_{ij}^{k}(t)$ matrices,
eq.~(\ref{rijb}),
with the help of the $\hat{\rho}_{00}^{k}(t)$ conditional 
environmental density matrix,
\begin{subequations}
	\label{rij}
\begin{eqnarray}
\label{r11}
\hat{\rho}_{11}^{k}(t)&=&\hat{w}^{k}(t)\hat{\rho}_{00}^{k}(t)\hat{w}^{k\dagger}(t),\\
\label{r01}
\hat{\rho}_{01}^{k}(t)&=&\hat{\rho}_{00}^{k}(t)\hat{w}^{k\dagger}(t),\\
\label{r10}
\hat{\rho}_{10}^{k}(t)&=&\hat{w}^{k}(t)\hat{\rho}_{00}^{k}(t),
\end{eqnarray}
\end{subequations}
where 
\begin{equation}
\label{w}
\hat{w}^{k}(t)=\hat{w}_1^{k}(t)\hat{w}_0^{k\dagger}(t).
\end{equation}

We can write the density matrix of the $k$-th environment conditional on
the qubit state being zero at time $t$ in its eigenbasis
\begin{equation}
\label{eigen}
\hat{\rho}_{00}^{k}(t)=\sum_{n_k}c_{n_k}|n_k(t)\rangle\langle n_k(t)|
\end{equation}
($c_{n_k}$ do not depend on $t$ because $\hat{\rho}_{00}^k(t)$ is obtained
from $\hat\rho^k(0)$ by a unitary rotation).
The basis of the density matrix $\hat{\rho}_{00}^{k}(t)$
is time-dependent, but in what follows we will
not write the time-dependences explicitly, both for clarity
and to save space.
Equally well, one can perform the following analysis in the eigenbasis of $\hat\rho^k(0)$ or $\hat{\rho}_{11}^{k}(t)$ and we pick $\hat{\rho}_{00}^{k}(t)$ for definiteness
and convenience.

The orthogonality condition (\ref{warunek}) 
can be written in this notation
\begin{eqnarray}
\label{war2}
\hat{\rho}_{00}^{k}(t)\hat{\rho}_{11}^{k}(t)&=&
\hat{\rho}_{00}^{k}(t)\hat{w}^{k}(t)\hat{\rho}_{00}^{k}(t)\hat{w}^{k\dagger}(t)\\
\nonumber
&=&
\sum_{n_k,m_k}c_{n_k}c_{n_k}|n_k\rangle\langle n_k|
\hat{w}^{k}(t)|m_k\rangle\langle m_k|\hat{w}^{k\dagger}(t)\\
\nonumber
&=&0.
\end{eqnarray}
For (\ref{war2}) to be fulfilled, all of the elements of the matrix
$\hat{\rho}_{00}^{k}(t)\hat{\rho}_{11}^{k}(t)$
must be equal to zero, so for all $p_k$ and $q_k$,
\begin{eqnarray}
\nonumber
\langle p_k|\hat{\rho}_{00}^{k}\hat{\rho}_{11}^{k}|q_k\rangle&=&
\sum_{n_k,m_k}c_{n_k}c_{m_k}\langle p_k|n_k\rangle\langle n_k|\hat{w}^{k}
|m_k\rangle\langle m_k|\hat{w}^{k\dagger}|q_k\rangle\\
\label{war3}
&=&c_{p_k}
\sum_{m_k}c_{m_k}\langle p_k|\hat{w}^k
|m_k\rangle\langle m_k|\hat{w}^{k\dagger}|q_k\rangle=0.
\end{eqnarray}
Here the second line is obtained assuming that the states $|p_k\rangle$
and $|q_k\rangle$ are eigenstates of $\hat{\rho}_{00}^k(t)$.
In this case, let us look at the diagonal elements, $p_k=q_k$, for which we
must have
\begin{equation}
\label{war4}
\langle p_k|\hat{\rho}_{00}^k(t)\hat{\rho}_{11}^k(t)|p_k\rangle
=c_{p_k}
\sum_{m_k}c_{m_k}|\langle p_k|\hat{w}^k(t)
|m_k\rangle|^2=0,
\end{equation}
for all $p_k$.
Since $c_{p_k}\ge 0$, $c_{m_k}\ge 0$, and $|\langle p_k|\hat{w}^k
|m_k\rangle|^2\ge 0$, the only situation when (\ref{war4})
is fulfilled, is either when $c_{p_k}=0$ or when $c_{m_k}= 0$ or $\langle p_k|\hat{w}^k
|m_k\rangle= 0$ for all $m_k$. Note that this automatically implies that the
condition (\ref{war3}) for off-diagonal elements is also met.
The last relevant, if somewhat trivial, observation here is that for $m_k=p_k$ (the sum spans over all 
$\hat{\rho}_{00}(t)$ eigenstates) we get that either $c_{p_k}= 0$ or $\langle p_k|\hat{w}^k
|p_k\rangle= 0$, which must hold for any $p_k$.

%Let us now return to the study of the density matrix (\ref{sigman}).
If the orthogonality condition is to be met for environment $k$, there must exist 
separate subspaces in the Hilbert space of the environment
for $\hat{\rho}_{00}^k(t)$ and $\hat{\rho}_{11}^k(t)$, so there must
exist eigenvalues of the conditional matrix $\hat{\rho}_{00}^k(t)$
which are equal to zero, $c_{q_k}=0$, since $\hat{\rho}_{11}^k(t)$
is a density matrix and cannot have all diagonal elements 
equal to zero. In other words $\hat{\rho}_{00}^k(t)$ must have  a non-trivial
kernel.
We denote the states corresponding
to this kernel as $|q_k\rangle$.

Since $\hat{w}^k(t)$ is unitary, we can always write 
\begin{equation}
\label{wn}
\hat{w}^k(t)|n_k\rangle = \sum_{m_k} b_{m_k}|m_k\rangle=
b_{n_k}|n_k\rangle +\sum_{m_k\neq n_k} b_{m_k}|m_k\rangle,
\end{equation}
with $\sum_{m_k} |b_{m_k}|^2=1$.
As shown previously, orthogonality implies
that either $b_{n_k}=\langle n_k|\hat{w}^k(t)|n_k\rangle =0$, or $c_{n_k}=0$,
so for all $n_k\neq q_k$ ($c_{n_k}\neq 0$), we have $b_{n_k} =0$
and
\begin{equation}
\label{wn2}
\hat{w}^k(t)|n_k\rangle =\sum_{m_k\neq n_k} b_{m_k}|m_k\rangle=|n_{_k\perp}\rangle,
\end{equation}
where $|n_{k\perp}\rangle$ is some state orthogonal to $|n_k\rangle$.

From (\ref{r11}) we have
\begin{equation}
\label{rho11}
\hat{\rho}_{11}^k(t)=\sum_{n_k}c_{n_k}|n_{k\perp}\rangle\langle n_{k\perp}|,
\end{equation}
and the orthonormality condition yields
\begin{eqnarray}
\label{ort}
\hat{\rho}_{00}^k(t)\hat{\rho}_{11}^k(t)&=&
\sum_{n_k}c_{n_k}|n_k\rangle\langle n_k|
\sum_{m_k}c_{m_k}|m_{k\perp}\rangle\langle m_{k\perp}|\\
\nonumber
&=&\sum_{n_k\neq m_k}c_{n_k}c_{m_k}
\langle n_k|m_{k\perp}\rangle
|n_k\rangle\langle m_{k\perp}|=0.
\end{eqnarray}
Since the condition (\ref{ort}) means that all of the elements of
the matrix must be equal to zero, it is equivalent to statement
that for all $n_k,m_k\neq q_k$,
\begin{equation}\label{ort_support}
\langle n_k|\hat w^k(t)|m_k\rangle=\langle n_k|m_{k\perp}\rangle=0,
\end{equation}
which means that the states $|n_{k\perp}\rangle$ must be orthogonal
not only to $|n_k\rangle$, but also to all other states in the support of
$\hat{\rho}_{00}^k(t)$ (eigenstates with non-zero occupations). This is not in the 
least surprising, because it simply means that the operator
$\hat{w}^k(t)$ takes the eigenstates of $\hat{\rho}_{00}^k(t)$
into a different subspace, as it should. Moreover, the states $|n_{k\perp}\rangle$
are orthogonal with respect to each other, since
\begin{equation}
\langle n_{k\perp}|m_{k\perp}\rangle=
\langle n_k|\hat{w}^{k\dagger}(t)\hat{w}^k(t)|m_k\rangle=
\langle n_k|m_k\rangle=\delta_{n_km_k}.
\end{equation}
Hence, they constitute a basis in the $\hat{\rho}_{11}^k(t)$
subspace and belong to the kernel of $\hat{\rho}_{00}^k(t)$
(our $|q_k\rangle$ states, for which
$c_{q_k}=0$). There can of course exist other states in $\mathrm{Ker}\hat{\rho}_{00}^k(t)$ but they play no role in our analysis.
%in the Hilbert space, all
%elements of the density matrix (\ref{sigman}) corresponding to these
%states are zero, and are irrelevant for the discussion).
Furthermore, the number of $|n_{k\perp}\rangle$ states has to be the same
as the number of $|n_k\rangle$ states, so the conditional density matrices
of the environment $\hat{\rho}_{ii}^k(t)$ are symmetric with respect
to each other, in the sense, that they have the same occupations
$c_{n_k}$, but for a different set of orthogonal eigenstates in different
subspaces, which immediately follows from (\ref{r11}). This further implies that for an environment of dimension $d_k$,
the dimension of the support of $\hat{\rho}_{ii}(t)$ cannot exceed $d_k/2$
for even $d_k$ and $(d_k-1)/2$ for odd $d_k$. 

The above analysis can be compactly summarized by writing matrix elements of $w^k(t)$ in the basis
$\{|r_k\rangle \} = \{|n_k\rangle\} \cup \{w^k(t)|m_k\rangle \}\cup \{|\kappa_k\rangle \}$, where the last vectors correspond to the part of the kernel of $\hat{\rho}_{00}^k(t)$ which is not of the form $w^k(t)|m_k\rangle$. Strict orthogonality at time $t$ implies that in such a chosen and ordered basis, we have the following matrix structure,
\begin{equation}\label{w_matrix}
\langle r_k|w^k(t) | r_k'\rangle =\left(
\begin{array}{ccc|ccc|c}  &  & & & & & \\
 & \mathbf 0 & & &\mathbf * & & \mathbf 0\\
  &  & & & & & \\ \hline
   &  & & & & & \\
    & \mathbf 1 & & & \mathbf 0 & & \mathbf 0\\
    &  & & & & & \\ \hline
    & \mathbf 0 & & & \mathbf * & & \mathbf *\\
\end{array} \right)
\end{equation}
In reality, strict orthogonality at time $t$ is a strong constraint on the free parameters of the model: The conditional evolutions 
$\hat V^i_k$, the initial environment state $\hat\rho^k(0)$, and the time moment $t$. Indeed, coming back the definitions (\ref{wib},\ref{rijb})
we have that $|n_k(t)\rangle = w_0^k(t) |n_k(0)\rangle$, where $|n_k(0)\rangle$ is the eigenbasis of 
$\hat\rho^k(0)$. Defining a new basis in the environment Hilbert space by $|r_k(0)\rangle = w_0(t)^\dagger |r_k(t)\rangle$,
we have that
\begin{equation}
\langle r_k(t)|w^k(t) | r'_k(t)\rangle = \langle r_k(0)|e^{i\hat{V}^0_kt}e^{-i\hat{V}^1_kt} | r'_k(0)\rangle
\end{equation}
and this matrix must have form (\ref{w_matrix}) at the time $t$.
%\tred{Zmienilam ci tu notacje, bo byla dla mnie niejasna. Jesli ci sie nie podoba to mozna 
%odwrocic. Przyznaje sie ze nie rozumiem tej czesci z macierza...}

\subsection{Purity}
Quite surprisingly, the 
emergence of objectivity puts some constraint on the initial purity of 
the environment.
%Since purity is conserved during a unitary evolution, we will be
%ale to find a lower bound on the initial purity of the environment
%necessary for the emergence of objectivity, when a single observer
%has access to only one environment. 
The constraint comes from the
orthogonality for the conditional evolution of each environmental state $\hat{\rho}_{ii}^k(t)$.
It is straightforward to show that the initial purity of the $k$-th environment
\begin{equation}
\label{purity}
P^k(0)=\Tr\hat{\rho}^{k}(0)^2
=\Tr\hat{\rho}_{00}^{k}(t)^2=\Tr\hat{\rho}_{11}^{k}(t)^2,
\end{equation}
using the definitions (\ref{rijb}), the fact that the operators
$\hat{w}_{i}^k(t)$ are unitary, and the properties of the trace.

In general the purity of a normalized state is bounded from below 
by $1/d_k$ (where $d_k$ is the dimension),
but we have previously shown that the dimensionality of the support of the
states $\hat{\rho}_{ii}^k(t)$ %in a single-environment-macrofraction  SBS state 
cannot exceed $d_k/2$ for 
even $d_k$ and $(d_k-1)/2$ for odd $d_k$.
%where $d$ is the size of the Hilbert space of the environment.
Hence, the minimal purity of $\hat{\rho}_{ii}^k(t)$ is $2/d_k$ or
$2/(d_k-1)$, respectively.
This means, as we see from eq.~(\ref{purity}), that %such an SBS state is only possible if the purity of the initial state
orthogonality is possible only when 
\begin{equation}
P^k(0) \geq \frac{2}{d_k}\ \text{ or }\  \frac{2}{d_k-1}.
\end{equation}
This doubly exceeds 
the minimum value for a mixed state purity and has non-negligible
consequences, especially for small environments. In the extreme case,
when each environment consists only of a qubit, it means that the emergence of objectivity
is only possible if the whole environment is initially pure. The same 
conclusion is drawn for qutrit environments.

\subsection{Consequences for qubit-environment entanglement}
Let us examine the consequences of strict orthogonality
for QEE. To this end, we will study the full system density matrix (\ref{caly})
without tracing out the degrees of freedom of the unobserved environment.
We do so since the resulting lack of coherence would conceal the level of qubit-environment 
entanglement necessary for the emergence of strict orthogonality
without any advantage to our understanding of the studied processes.
This matrix can be written at time $t$ using the eigenstates of the unobserved 
environmental matrix $\hat{R}_{00}(t)$ and each unobserved environmental matrix 
$\hat{\rho}_{00}^k(t)$ at time $t$, $\{|r\rangle\}$ and $\{|n_k\rangle\}$, 
respectively.
To do this we introduce the joint matrices describing the evolution
of the observed environments
$\hat{\rho}_{ij}(t)=\bigotimes_k\hat{\rho}_{ij}^k(t)$
and the joint conditional evolution operator of the observed environment
$\hat{w}(t)=\bigotimes_k\hat{w}^k(t)$.
Using equations analogous to eqs (\ref{rij}) we obtain
\begin{widetext}
\begin{equation}
\label{sigma}
\hat{\tilde{\sigma}}(t)=\left(
\begin{array}{cc}
|a_0|^2\hat{R}_{00}(t)\otimes\hat{\rho}_{00}(t)&
a_0a_1^*(t)\hat{R}_{00}(t)\hat{w}^{not\dagger}(t)\otimes\hat{\rho}_{00}(t)\hat{w}^{\dagger}(t)\\
a_0^*a_1(t)\hat{w}^{not}(t)\hat{R}_{00}(t)\otimes\hat{w}(t)\hat{\rho}_{00}(t)&
|a_1|^2\hat{w}^{not}(t)\hat{R}_{00}(t)\hat{w}^{not\dagger}(t)\otimes\hat{w}(t)\hat{\rho}_{00}(t)\hat{w}^{\dagger}(t)
\end{array}
\right),
\end{equation}
\end{widetext}
where $\hat{w}^{not}(t)=\hat{w}_1^{not}(t)\hat{w}_0^{not\dagger}(t)$.
We now
introduce eigenstates of the conditional density matrix of all observed environments
\begin{equation}
\label{m}
|n\rangle =\bigotimes_k|n_k\rangle =|n_1\dots n_N\rangle,
\end{equation}
where $N$ is the number of such environments,
so the their joint conditional density matrix must be of the form
\begin{equation}
\label{duzerho}
\hat{\rho}_{00}(t)=\sum_n C_n|n\rangle\langle n|,
\end{equation}
where the eigenvalues corresponding to each eigenstate $|n\rangle$ are (cf.~(\ref{eigen}))
\begin{equation}
C_n=\prod_{k=1}^{N} c_{n_k},
\end{equation}
The summation in eq.~(\ref{duzerho}) is over all possible values of $n$, so it is over all possible combinations of 
the eigenstates $|n_k\rangle$ for different observed environments.
We can similarly decompose the conditional density matrix of the unobserved 
environment,  
\begin{equation}
\label{malerho}
\hat{R}_{0}(t)=\sum_r c_r|r\rangle\langle r|,
\end{equation}
where $c_r$ and $|r\rangle$ 
are its eigenvalues and eigenvectors at time $t$, respectively.

Inserting eqs (\ref{duzerho}) and (\ref{malerho}) into eq.~(\ref{sigma}) yields
\begin{equation}
\label{sigman}
\hat{\tilde{\sigma}}=
\sum_{r,n}c_rC_n|\psi_{r,n}\rangle\langle\psi_{r,n}|,
\end{equation}
where
\begin{equation}
\label{psin}
|\psi_{r,n}\rangle =a_0|0\rangle\otimes|r\rangle\otimes|n\rangle+a_1(t)|1\rangle\otimes
\hat{w}^{not}(t)|r\rangle\otimes\hat{w}(t)|n\rangle.
\end{equation}
Such a decomposition into projectors is possible
only because qubit is initially in a pure state; if this was not the case an analogous 
decompositon could be made into density matrices.

The orthogonality condition (\ref{ort_support}) for environment $k$ implies that for all 
states $|n_k\rangle$, $|m_k\rangle$ in the support of $\hat{\rho}_{00}^k(t)$,
$\langle m_k|n_{k\perp}\rangle=\langle m_k|\hat{w}^k(t)|n_k\rangle=0$. 
%is orthogonal not only to $|n_k\rangle$,
%but also to any $|m_k\rangle$. 
This automatically translates into maximum possible 
entanglement for a given initial qubit occupation $a$
between the qubit and the rest of the system 
in a state $|\psi_{r,n}\rangle$, since the states $|r\rangle\otimes |n\rangle$ and
$\hat w^{not}(t)|r\rangle \otimes \hat w(t)|n\rangle$ are orthogonal.
In fact, the state has maximum possible entanglement 
between any subspace containing the qubit and any subspace containing the environment 
that fulfills the orthogonality condition.

Another consequence of the orthogonality condition being fulflled for even one environment
is the guaranteed full dephasing of the qubit due to the qubit-environment interaction. Tracing
out over the environmental degrees of freedom yields
\begin{equation}
\Tr_E\hat{\sigma}(t)=|a_0|^2|0\rangle\langle 0|
+|a_1|^2|1\rangle\langle 1|,
\end{equation}
because the orthogonality condition (\ref{ort_support}) for environment $k$ gives 
$\mathrm{Tr}[\hat{w}_{i}^k(t)\hat{\rho}_{00}^k(t)]=0$
which kills the off-diagonal elements in (\ref{sigma}) %the density matrix of the qubit,
regardless of the other envirnments.
This is a manifestation of a fact that if considered in the same subspace, orthogonality is a stronger condition than decoherence. Indeed one can easily show that the modulus of the decoherence factor for the $k$-th environment
is not greater than the generalized overlap between the states $\hat{\rho}_{00}^k(t)$ and $\hat{\rho}_{11}^k(t)$:
\begin{equation}
\left|\Tr\left[\hat{w}_1^{k\dagger}(t)\hat{w}_0(t)\hat{\rho}^k(0)\right]\right|\leq \mathrm{Tr}\left[\sqrt{\hat{\rho}_{00}^k(t)}
\hat{\rho}_{00}^k(t)\sqrt{\hat{\rho}_{00}^k(t)}\right]^{\tfrac{1}{2}}.
\end{equation}

For an SBS state to emerge, strict orthogonality must be fulfilled
for all observed environments (for all $k$). %and we have
%$|n_{\perp}\rangle=\hat{w}(t)|n\rangle$ which must be perpendicular
%to any state outside of the kernel of $\hat{\rho}_{00}(t)$, $|m\rangle$.
As is straightforward to see, all of the states
(\ref{psin}) which enter the decomposition (\ref{sigman})
must be defined on different subspaces of the full Hilbert space, so
it is straightforward to find 
entanglement (as measured by Negativity $\mathcal N$) between the qubit
and all of the environment. 
This is because the qubit-environment density matrix (\ref{sigman})
is in this case block-diagonal in subspaces of 
$\{|0\rangle\otimes|r\rangle\otimes|n\rangle,
|1\rangle\otimes|r\rangle\otimes|n\rangle,
|0\rangle\otimes|r\rangle\otimes|n_{\perp}\rangle,
|1\rangle\otimes|r\rangle\otimes|n_{\perp}\rangle\}$,
for a given $|n\rangle$ and $|n_{\perp}\rangle$ from the support of $\hat{\rho}_{00}^k(t)$ and an arbitrary
$|r\rangle$.
This means that 
the block diagonal form persists after partial transposition. Since Negativity \cite{vidal02,lee00a} is the 
absolute value of the sum of all negative eigenvalues of a density 
matrix after partial transposition with respect to one of the (potentially) entangled subsystems, for such block diagonal matrices,
negativity will be the sum of the Negativities of 
$|\psi_{r,n}\rangle\langle \psi_{r,n}|$ weighted by the coefficients $c_rC_n$, as $|\psi_{r,n}\rangle$ are all orthogonal to each other.

The Negativity of each matrix $|\psi_{r,n}\rangle\langle \psi_{r,n}|$ is
the same and
is equal to $|a_0a_1|$, cf. (\ref{psin}), so the Negativity of 
the density matrix (\ref{sigman}) is also equal to $|a_0a_1|$
\begin{equation}
\mathcal N (\hat{\tilde{\sigma}}(t))=|a_0a_1|,
\end{equation}
since $\sum_r c_r\sum_n C_n =1$.
This is fairly surprising, since
the value of Negativity which allows for the formation of SBS states
is set only by the occupation of qubit pointer states, and does not
depend on the dimension of the observed environments. It also does not
depend on the purity of the state, as long as the purity is high
enough that it allows for the formation of states which fulfill 
the orthonormality criterion (see previous subsection).
Note, that 
this is the same value of Negativity that one would have, if the environment
would be a qubit. In this case the qubit-environment state has
to be pure (again see previous subsection) and the qubit-environment
state in question has the largest possible amount of entanglement
for a given initial qubit superposition state.

An important observation here is that if the value of Negativity, $|a_0a_1|$, between the qubit
and its environments is reached, this does not automatically mean that all of the
observed environments fulfill the strict orthogonality criterion. It only means that
at least one of the environments does. Hence, to probe orthogonality
via the created entanglement in a
system with many observed
environments, it would be necessary to find the Negativity between each observed environment
and the rest of the system separately. The fulfillment of the strict objectivity criterion
would be accompanied by the maximization of such Negativity, when it would reach
the value of $|a_0a_1|$.  

\section{Conclusions\label{sec5}}
We have studied the situation when a system interacts with  multiple observed
and unobserved environments 
to determine the interdependencies
between system-environment entanglement generation and the possibility of the
emergence of objectivity via SBS states (\ref{SBS_0}). Such entanglement does not change the 
properties of decoherence  \cite{Eisert_PRL02,Hilt_PRA09,Pernice_PRA11,roszak15a,roszak18} and hence is irrelevant for the
decoherence function, so the
effect of the unobserved environments is independent of
qubit-environment entanglement formation. On the other hand, 
the entanglement turns out to be crucial when it comes to
the ability of the observed environments to store information 
about the state of the decohering qubit.

Firstly, we have shown that the lack of system-environment entanglement
generation is synonymous with the imposibility of an SBS state formation
with respect to the pointer basis of the qubit. This is true regardless of the
number of environments in each observers' macrofraction, since
separability translates into the environmental states conditioned on
the qubit state being completely indistinguishable.

For a qubit system, we have further studied case of ideal distinguishability, 
when the supports of the 
environmental states conditioned on
the qubit state are strictly orthogonal. It turns out that such orthogonality
requires the qubit-observed-environment state
to take an extreme entangled form, which is characterized by the same amount
of Negativity as would be found in the pure state $a_0|00\rangle +a_1|11\rangle$ ($\mathcal N =|a_0a_1|$) for an environment consisting
of a single qubit. Additionally, we have found that such states
cannot appear if the initial qubit-observed-environment state
is of too low a purity.
In fact, the limitations on the initial purity of the environment
are quite strong (the purity must be twice as big as the minimum possible
purity of a system of a given size), and are especially important for 
small environments. This limitation 
in the extreme case of qubit or qutrit environments
leads to the requirement of a pure initial state of all environments.
Otherwise the emergence of objectivity (SBS) is impossible.

It is known that in many situations 
\cite{tuziemski15,mironowicz17,korbicz17,lampo17,korbicz17a,tuziemski18,mironowicz18,tuziemski19},
a single environment being observed by a single observer is insufficient
for the emergence of objectivity, but the addition of more environments
to the observers disposal (forming a so called macrofraction) can lead to SBS states in the limit of a large
number of environments. Since we have here qualified the correspondence
between qubit-environment entanglement in the extreme cases
and shown that separability excludes
objectivity, while the natural emergence of SBS states for 
an observer limited to a single environment requires the
qubit and the observed environment to entangle strongly,
we conjecture that the amount of the generated QEE manifests itself in the way that the SBS state
is approached with the growing number of environments in a single macrofraction. Since weakly entangled systems
feature weakly distinguishable conditional environmental states,
it follows that the more qubit-environment entanglement present
in the system, the faster a near-SBS state should be approached with
growing macrofractions. Consequently, not only is entanglement a
necessary condition for objectivity, but objectivity is to be expected
more readily in systems which strongly entangle during their joint
evolutions with their environments.

\section*{Acknowledgements}
The authors would like to thank {\L}ukasz Cywi{\'n}ski for helpful discussions about qubit decoherence. 
This work is supported by funds from Polish National Science
Center (NCN) Grant No. DEC-2015/19/B/ST3/03152.

\end{document}